\documentstyle[psfig,12pt,twoside,fleqn,espcrc1]{article}

\input epsf.sty

\def\ergs{ergs s$^{-1}$ }
\def\gmc{g cm$^{-3}$}
\def\kms{km s$^{-1}$}
\def\ms{$M_\odot$}
\def\ni{$^{56}$Ni}
\def\e#1{$\times$ $10^{#1}$ }
\def\ee#1{$10^{#1}$ }
\def\ltsima{$\; \buildrel < \over \sim \;$}
\def\ltsim{\lower.5ex\hbox{\ltsima}}
\def\gtsima{$\; \buildrel > \over \sim \;$}
\def\gtsim{\lower.5ex\hbox{\gtsima}}
\def\etal{{\sl et al.} }

\title{NUCLEOSYNTHESIS IN TYPE IA SUPERNOVAE}

\author{K. Nomoto{\hbox{$^{\rm a}$}}, K. Iwamoto{\hbox{$^{\rm a}$}},
N. Nakasato\address{Department of Astronomy, University of Tokyo,Tokyo 113},
F.-K. Thielemann{\hbox{$^{\rm b}$}}, \\ F. Brachwitz\address{Institut f\"ur
Theoretische Physik, Universit\"at Basel CH-4056 Basel, Switzerland},
T. Tsujimoto\address{National Astronomical Observatory, Mitaka, 
Tokyo 181}, Y. Kubo{\hbox{$^{\rm a}$}}, N. Kishimoto{\hbox{$^{\rm a}$}}
}

\begin{document}
\maketitle

\begin{abstract}

     Among the major uncertainties involved in the Chandrasekhar mass
models for Type Ia supernovae are the companion star of the accreting
white dwarf (or the accretion rate that determines the carbon ignition
density) and the flame speed after ignition.  We present
nucleosynthesis results from relatively slow deflagration (1.5 - 3 \%
of the sound speed) to constrain the rate of accretion from the
companion star.  Because of electron capture, a significant amount of
neutron-rich species such as $^{54}$Cr, $^{50}$Ti, $^{58}$Fe,
$^{62}$Ni, etc. are synthesized in the central region.  To avoid the
too large ratios of $^{54}$Cr/$^{56}$Fe and $^{50}$Ti/$^{56}$Fe, the
central density of the white dwarf at thermonuclear runaway must be as
low as \ltsim 2 \e9 \gmc.  Such a low central density can be realized
by the accretion as fast as $\dot M$ \gtsim 1 $\times$ 10$^{-7}
M_\odot$ yr$^{-1}$.  These rapidly accreting white dwarfs might
correspond to the super-soft X-ray sources.

\end{abstract}

\section {INTRODUCTION}

     Supernovae of different types have different progenitors, thus
producing different heavy elements on different time scales during the
chemical evolution of galaxies.  Because the lifetime of their massive
progenitors is about $10^{6-7}$ years being much shorter than the age
of galaxies, Type II supernovae (SNe II) and Type Ib/Ic supernovae
(SNe Ib/Ic) cause the heavy-element enrichment in the early phase of
the galactic evolution.  

     In contrast, Type Ia supernovae (SNe Ia) produce heavy elements on
a much longer time scale in the later phase of the galactic evolution.
There are strong observational and theoretical indications that SNe Ia
are thermonuclear explosions of accreting white dwarfs (e.g., Nomoto
\etal 1994 for a review).  However, the exact binary evolution that
leads to SNe Ia has not been identified (e.g., Renzini 1996; Branch
\etal 1995 for recent reviews).  The identification of the
progenitor's evolution is critically important 1) for clarifying
whether the nature of SNe Ia at high redshift is the same as nearby
SNe Ia, and 2) for understanding the origin of some systematic
variations of light curves (i.e., brighter--slower), which is a
critical material for the determination of cosmological parameters,
$H_0$ and $q_0$.  In this paper, we provide some important  
constraints on the progenitor system from the viewpoint of  
nucleosynthesis, namely, the carbon ignition density which is 
translated into the accretion rate for the Chandrasekhar mass models.  

\section {NUCLEOSYNTHESIS IN SLOW DEFLAGRATION}

\begin{figure}
\epsfxsize=300pt
\epsfbox{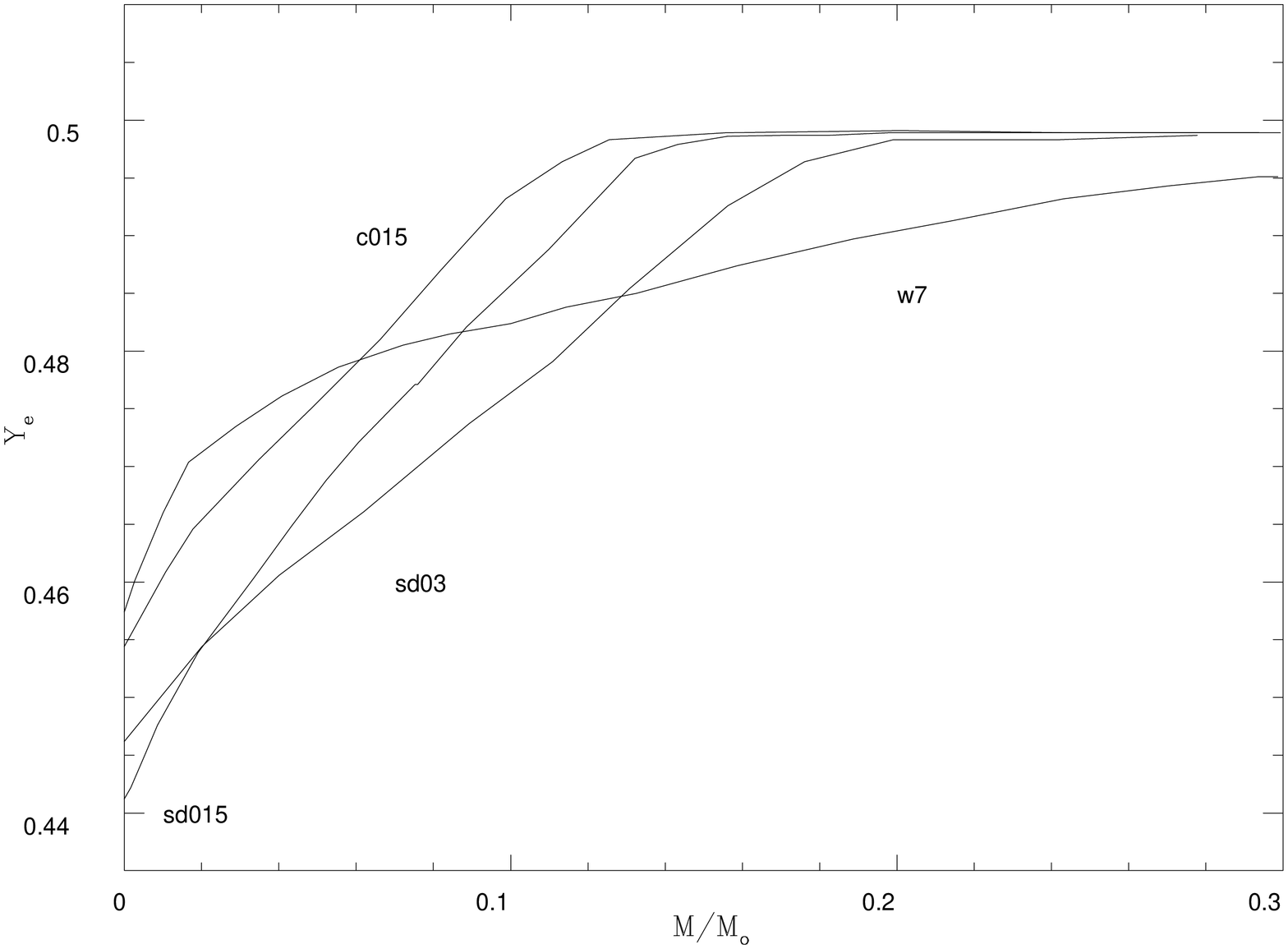}
\caption{$Y_{\rm e}$, the total proton to nucleon ratio, and thus a
measure of electron captures on proton and nuclei, as a function of
radial mass for different models.  Here c015, sd015, and sd030
correspond to CS15, WS15, and WD30 in the text.}
\label{tnib5}
\end{figure}

     For the Chandrasekhar mass white dwarf model, carbon burning in
the central region leads to a thermonuclear runway.  A flame front
then propagates at a subsonic speed as a {\sl deflagration wave} due
to heat transport across the front (e.g., Nomoto \etal 1996a for a review). 
Here the major uncertainty is the flame speed which depends on  the 
development of instabilities of various scales at the flame front.        
Multi-dimensional hydrodynamical simulations of the flame propagation 
have been attempted by several groups (Livne 1993; Arnett \& Livne 1994; 
Khokhlov 1995; Niemeyer \& Hillebrandt 1995).  These simulations have 
suggested that a carbon deflagration wave propagates at a speed 
$v_{\rm def}$ as slow as 1.5 - 3 \% of the sound speed $v_{\rm s}$ in 
the central region of the white dwarf.  Though the calculated flame speed 
is still very preliminary, it is useful to examine nucleosynthesis 
consequences of such a slow flame speed.

     In the deflagration wave, electron capture enhances neutron  
excess, which depends on both $v_{\rm def}$ and the central density of  
the white dwarf $\rho_9 = \rho_{\rm c}$/10$^9$ g cm$^{-3}$.  The  
resultant nucleosynthesis in slow deflagration has some distinct  
features compared with the faster deflagration like W7
(Nomoto \etal 1984; Thielemann \etal 1986), thus providing  
important constraints on these two parameters.  The constraint on the  
central density is equivalent to a constraint on the accretion rate.  

     We calculate explosive nucleosynthesis for two cases with
$\rho_9$ = 1.37 (C) and 2.12 (W) at the thermonuclear runaway, i.e.,
at the stage when the timescale of the temperature rise in the center
becomes shorter than the dynamical timescale.  Here C and W imply that
the models are the same as calculated for C6 and W7, respectively
(Nomoto \etal 1984).  For the slow (S) deflagration, we adopt $v_{\rm
def}/v_{\rm s}$ = 0.015 (WS15, CS15) and 0.03 (WS30).  The central
region behind the slower deflagration undergoes electron capture for a
longer time than in W7, thereby having significantly reduced $Y_{\rm
e}$.  In Figure \ref{tnib5}, profiles of $Y_{\rm e}$ for these cases
and W7 are shown (Thielemann \etal 1996b). In general it can be recognized 
that small burning front velocities lead to steep $Y_{\rm e}$-gradient 
which flatten with increasing velocities.  Lower central ignition densities 
shift the curves up (CS15), but the gradient is the same for the same
propagation speed.

\begin{figure}[t]
\epsfxsize=300pt
\epsfbox{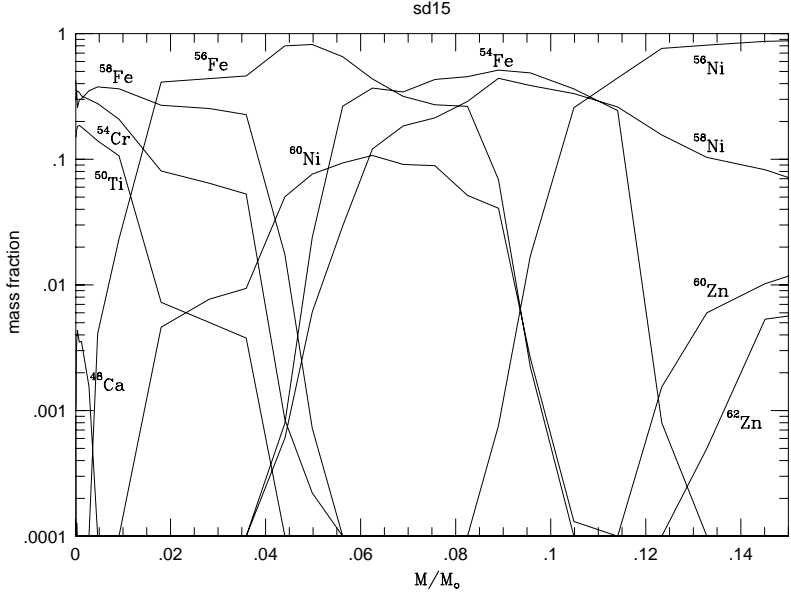}
\caption{The abundance distribution in slow deflagration WS15.}
\label{tnib6}
\end{figure}

     Figure \ref{tnib6} shows the abundance distribution of
neutron-rich species such as $^{54}$Cr, $^{50}$Ti, $^{58}$Fe, and
$^{62}$Ni behind the slow deflagration for WS15.  The locations of
$^{54}$Fe and $^{58}$Ni, overproduced in W7, correspond to $Y_{\rm e}$
= 0.47-0.485.  Due to the $Y_{\rm e}$-gradients which are steeper than
for W7, the amount of matter in a given $Y_{\rm e}$-range is reduced,
but also smaller central values are attained, giving rise to more
neutron-rich nuclei.  $Y_{\rm e} \sim$ 0.46 $\approx$ 26/56 leads to a
large abundance of stable $^{56}$Fe (not from $^{56}$Ni decay).
$Y_{\rm e}$ = 0.44 - 0.46 result also in $^{48}$Ca, $^{50}$Ti,
$^{54}$Cr, and $^{58}$Fe.  Of these nuclei $^{48}$Ca with Z/A
$\approx$ 0.42 is only produced if $Y_{\rm e}$ approaches values close
to and smaller than 0.44 (Woosley 1996; Meyer \etal 1996).  As the
deflagration wave propagates outwards, the white dwarf gradually
expands to undergo less electron capture and thus mostly \ni~ is
synthesized.  Eventually, the deflagration enters the region of
incomplete Si burning and explosive O-Ne-C burning.

     The neutron excess is sensitive to the initial central density.
For WS15 ($\rho_9$ = 2.12), the masses of $^{54}$Cr and $^{50}$Ti are
smaller than those for CS15 ($\rho_9$ = 1.37) by a factor of $\sim$ 5
and 20, respectively, and those for WS30 ($v_{\rm def}/v_{\rm s}$ =
0.03) by a factor of 2 and 4, respectively.

\section {NUCLEOSYNTHESIS IN DELAYED DETONATION}

\begin{figure}
\epsfxsize=300pt
\epsfbox[37 384 557 752]{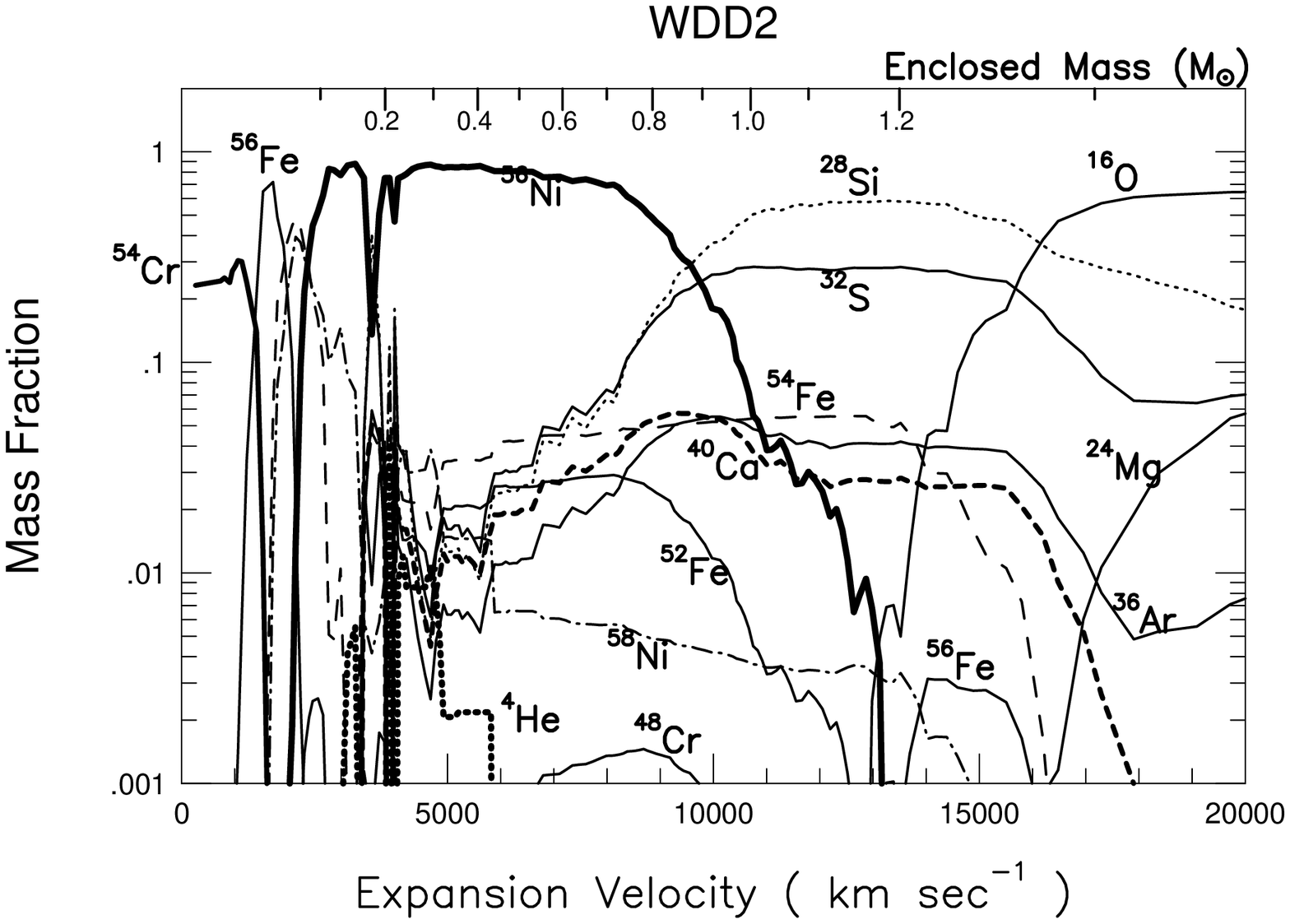}
\caption{The abundance distribution in the delayed detonation model  
WDD2 as a function of interior mass and the expansion
velocity.}
\label{wdd2}
\end{figure}

\begin{figure}[t]
\epsfxsize=300pt
\epsfbox[37 384 557 752]{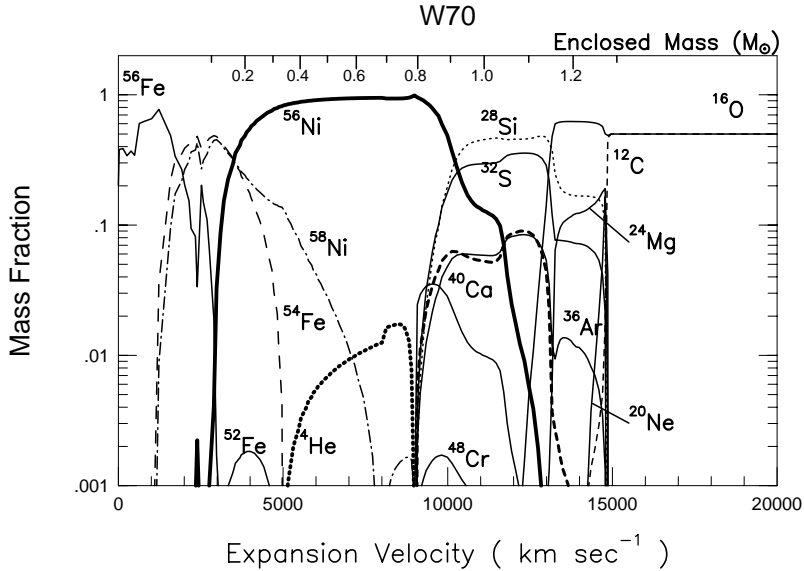}
\caption{The abundance distribution in the delayed detonation model  
W70 as a function of interior mass and the expansion
velocity.}
\label{w70}
\end{figure}

     If the deflagration speed continues to be much slower than in W7,  
the white dwarf undergoes strong pulsation as first found by Nomoto 
\etal (1976).  In this {\sl pulsating deflagration} model, the white 
dwarf expands to quench nuclear burning when the total energy of the 
star is still negative.  Then the star contracts to burn more material 
to make the total energy positive ($\sim$ 5 \e{49}erg s$^{-1}$).  

     The deflagration might induce a detonation at low density layers.
In the {\sl delayed detonation} model (Khokhlov 1991a; Woosley \&
Weaver 1994), the deflagration wave is assumed to be transformed into
detonation at a certain layer during the first expansion phase.  In
the {\sl pulsating} delayed detonation model (Khokhlov 1991b), the
transition into detonation is assumed to occur near the maximum
compression due to mixing.

     We study explosive nucleosynthesis with a large reaction network 
(Thielemann \etal 1996) for various delayed detonation (DD) models, 
which has not been discussed in detail before (Nomoto \etal 1996b).  We 
artificially transform the slow deflagration WS15 into detonation when 
the density ahead of the flame decreases to 3.0, 2.2, and 1.7 \e7 \gmc~ 
(WDD3, WDD2, and WDD1, respectively, where 3, 2, and 1 indicate $\rho_7$ at 
the transition).  Then the carbon detonation propagates through the 
layers with $\rho < $ \ee8 \gmc.  The explosion energy of three WDD
models is 1.5 \e{51} \ergs and the mass of \ni~ is 0.73 \ms~ (WDD3),
0.58 \ms~ (WDD2), and 0.45 \ms~ (WDD1).  

     Figures \ref{wdd2} shows the abundance distribution against the
expansion velocity and $M_r$ after the passage of the slow
deflagration (WS15) and the delayed detonation.  For comparison, the
abundance distribution of W70 is shown in Figure \ref{w70}, where the
initial metallicity (i.e., the initial CNO elements which are later
transformed into $^{22}$Ne) is assumed to be zero.  

     It is seen that WDD2 and WDD1 produce two Si-S-Ar peaks at low
velocity ($\sim$ 4000 \kms) and high velocities (10,000 - 15,000 km
s$^{-1}$).  The low velocity intermediate mass elements are important
to observe at late times to distinguish models.  In particular, the
minimum velocity of Ca in WDD models is $\sim$ 4000 \kms, which should
be compared with the observed minimum velocities of Ca indicated by
the red edge of the Ca II H and K absorption blend (Fisher \etal
1995).

     Meikle \etal (1996) have observed a P Cyg-like feature at $\sim$ 
1.05/1.08 $\mu$m in SN 1994D and 1991T.  They note that, if this 
feature is due to He, He in SN 1994D is likely to be formed in 
$\alpha$-rich freezeout and mixed out to the high velocity layers 
($\sim$ 12,000 \kms).  The maximum velocity of He is 5000 - 6000 \kms~ 
in WDDs being slower than $\sim$ 9000 \kms~ in W7, so that more 
extensive mixing of He would be required for WDDs than W7.  
Alternatively, if the feature is due to Mg, the Mg velocity is confined 
in 12,500 - 16,000 \kms~ in SN 1994D, which is consistent with W7 
(13,000 - 15,000 \kms).  For WDDs, on the other hand, the minimum 
velocities of Mg are 14,500 \kms~ (WDD1), 16,500 \kms~ (WDD2), and 
18,000 \kms~ (WDD3), and the latter two models seem to have too high 
velocities.  

\section {YIELDS OF TYPE IA SUPERNOVAE}

\begin{figure}
\epsfxsize=300pt
\epsfbox{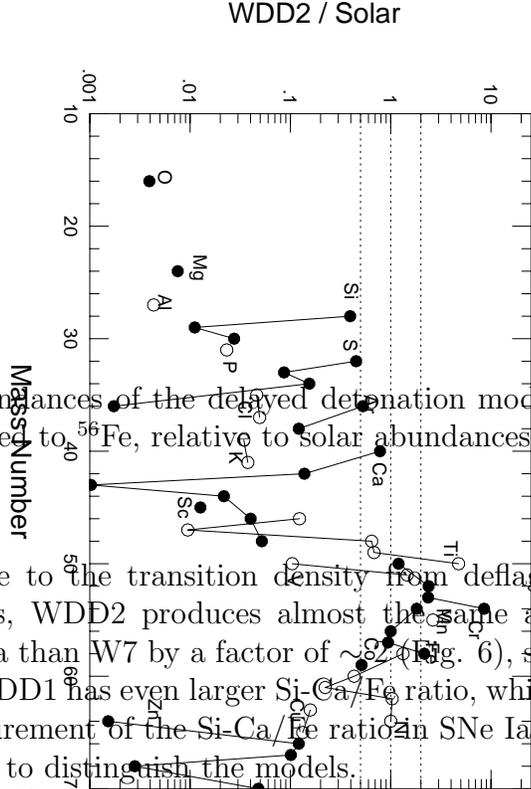}
\caption{The ratios of integrated abundances of the delayed detonation 
model WDD2 after decay of unstable nuclei, normalized to $^{56}$Fe,  
relative to solar abundances.}
\label{wdd2sum}
\end{figure}

\begin{figure}[p]
\epsfxsize=300pt
\epsfbox{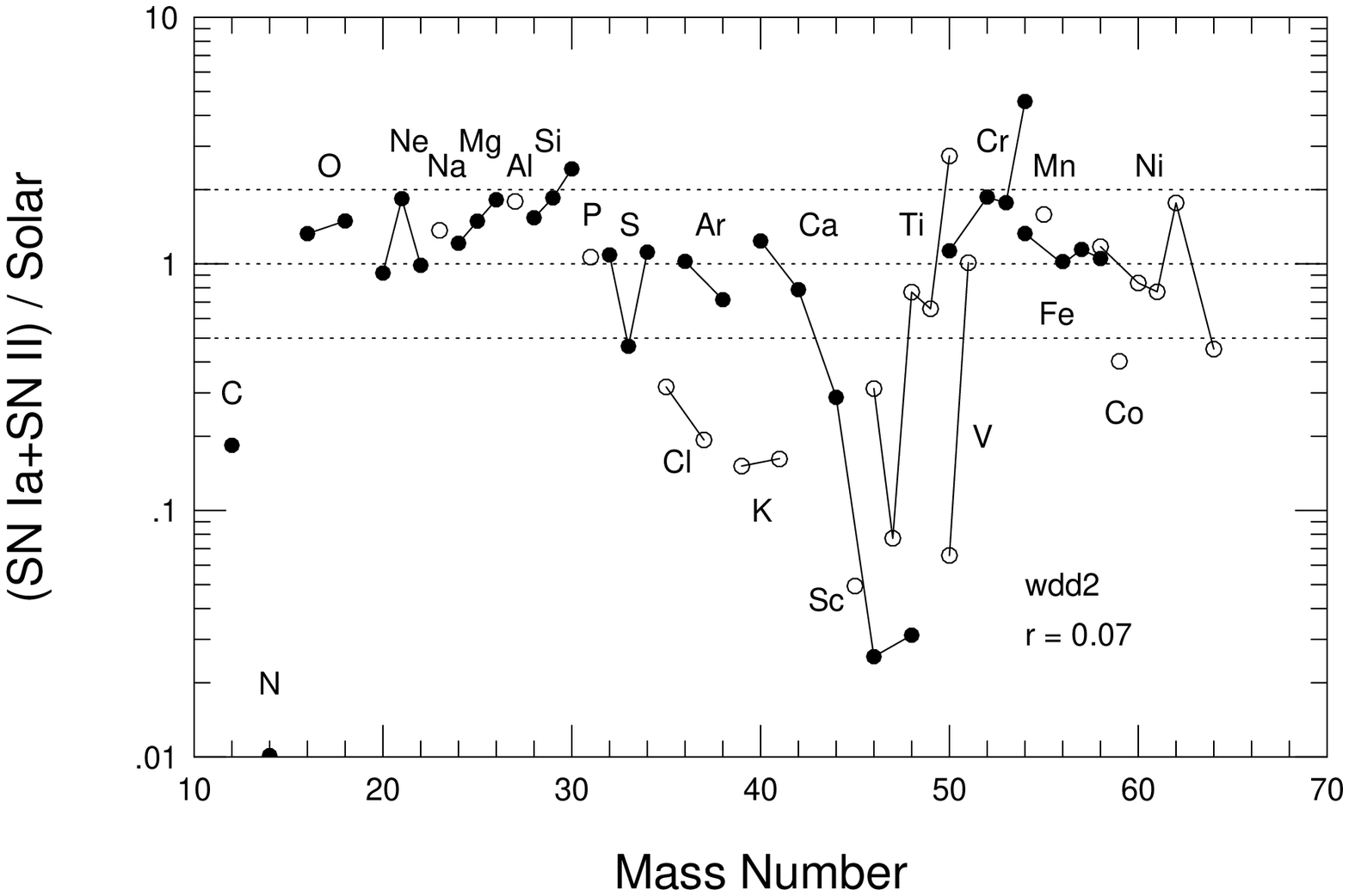}
\caption{
Solar abundance pattern based on synthesized heavy elements from a 
composite of Type Ia and Type II supernova explosions with the most 
probable ratio of $r$.  Relative abundances of synthesized heavy 
elements and their isotopes normalized to the corresponding solar 
abundances are shown by circles.  Here WDD2 is adopted for the
Type Ia supernova model.}
\label{wdd2sol}
\vspace{20mm}
\epsfxsize=300pt
\epsfbox{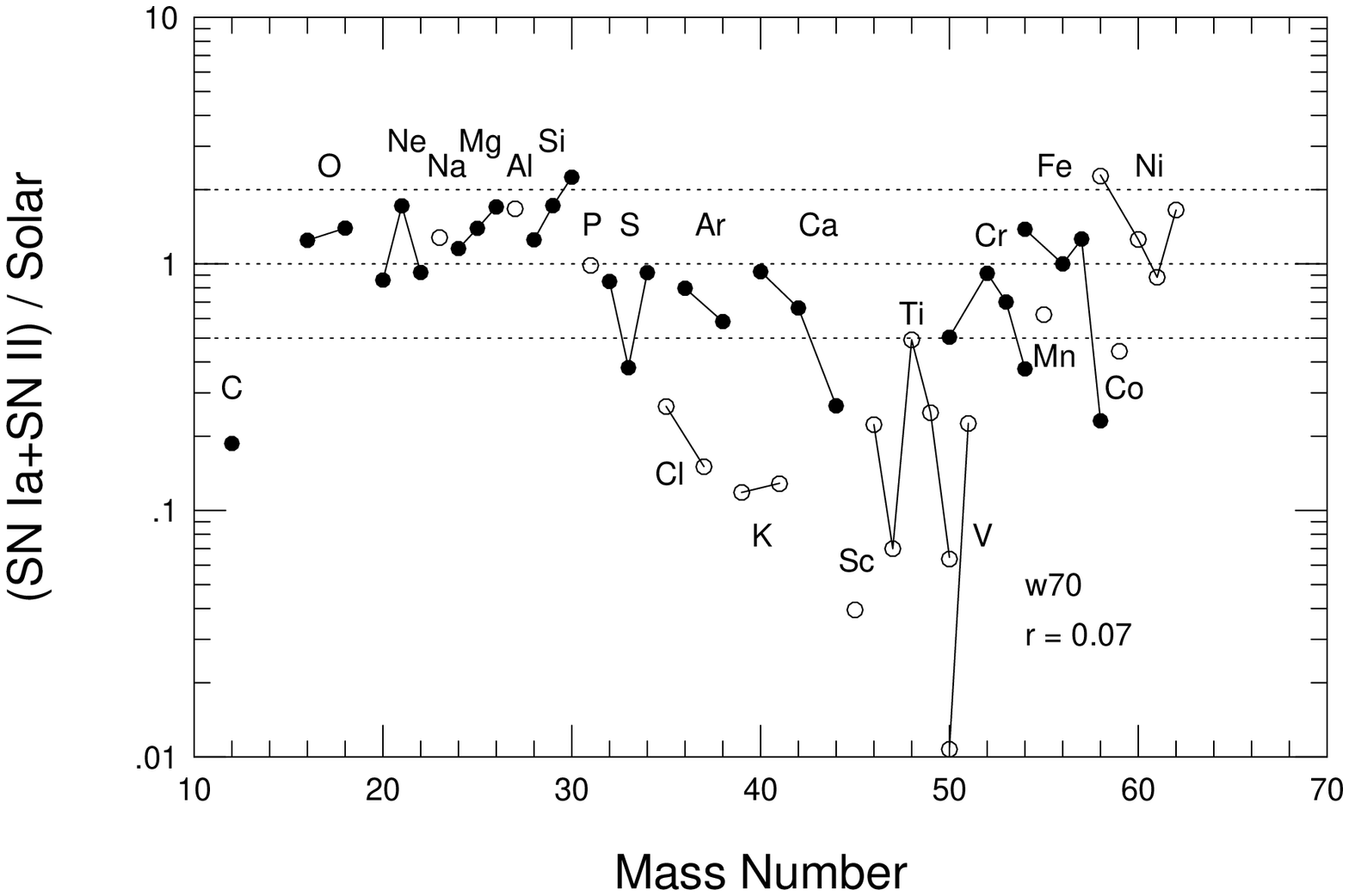}
\caption{
Same as Figure \protect\ref{wdd2sol} but for W70 as an SN Ia model.}
\label{w70sol}
\end{figure}

     Total isotopic compositions of WDD1 - WDD3 are given in Table
\ref{tabmas}, and compared with the solar abundances in Figure
\ref{wdd2sum} which are normalized to $^{56}$Fe.  Table \ref{tabmas}
also includes W7 and W70 updated with the latest reaction rates along
with W70 (Thielemann \etal 1996b).  We note:

1) The synthesized amounts of Fe and thus the ratio between the
intermediate mass elements to Fe, Si-Ca/Fe, are sensitive to the
transition density from deflagration to detonation.  Among the WDD
models, WDD2 produces almost the same amount of \ni~ as W7 ($\sim$ 0.6
\ms) but more Si-Ca than W7 by a factor of $\sim$ 2 (Fig.
\ref{wdd2sol}), since more oxygen is burned in the outer layers.  WDD1
has even larger Si-Ca/Fe ratio, which is close to the solar ratio.
Therefore, the measurement of the Si-Ca/Fe ratio in SNe Ia remnants
(Tycho, SN 1006, etc.) would be useful to distinguish the models.

2) Some neutron-rich species such as $^{54}$Cr and $^{50}$Ti are
overproduced with respect to $^{56}$Fe.  To see the degree of
overproduction, we combine nucleosynthesis products of SNe Ia and SNe
II with various ratios and compare with solar abundances of heavy
elements and their isotopes.

     Nucleosynthesis products of SNe II as a function of stellar
masses are taken from the calculations by Nomoto \& Hashimoto (1988),
Hashimoto \etal (1989, 1996) and Thielemann \etal (1996a) as summarized
in Tsujimoto \etal (1995).  SNe II yields integrated over $m_l$ = 10
\ms~ to $m_u$ = 50 \ms~ with the Salpeter IMF are given in Table
\ref{tabmas}.  The upper mass bound $m_u$ is chosen to give [O/Fe] =
+0.4, which is consistent with the observations of low metallicity
stars for [Fe/H] $<$ --1.

     Nucleosynthesis products of SNe Ia are those from WDD2 and W70,
and the results are shown in Figures \ref{wdd2sol} - \ref{w70sol}.
For WDD2, the best fit to the solar abundances are obtained for the
ratio between SNe Ia and SNe II contributions as $r$ = 0.07 where $r$
is the mass fraction contributed by SNe Ia per unit mass of all heavy
elements in the gas as defined in Tsujimoto \etal (1995).

     Aided with a reasonable chemical evolution model (Yoshii \etal
1996), the number of SNe Ia ever occurred relative to SNe II is 
determined to be $N_{\rm Ia}/N_{\rm II}=0.12$ in order to reproduce the 
observed abundances.  This is consistent with the fact that the 
observed estimate of SNe Ia frequency is as low as 10 \% of the total 
supernova occurrence (van den Bergh \& Tammann 1991; Tsujimoto \etal 
1995).

     With this relative frequency, $^{56}$Fe from SNe Ia is about 50
\% of total $^{56}$Fe.  Then the abundance ratios between neutron-rich
species and $^{56}$Fe could be reduced by a factor of $\sim$ 2.  For
W70, the excess of $^{58}$Ni/$^{56}$Fe is now within the uncertainties
(Fig. \ref{w70sol}).  For WDDs, on the other hand, $^{54}$Cr and
$^{50}$Ti are overproduced as seen in Figure \ref{wdd2sol} even when
the contribution from SNe II is taken into account.

     The above results imply that the central density of the 
Chandrasekhar mass white dwarf at thermonuclear runaway must be as low 
as \ltsim 2 \e9 \gmc, though the exact constraint depends somewhat on 
the flame speed.  Such a low central density can be realized by the 
accretion as fast as $\dot M >$ 1 $\times$ 10$^{-7} M_\odot$ yr$^{-1}$.  
Such rapidly  accreting white dwarfs might correspond to the super-soft 
X-ray  sources.

\section*{Acknowledgements}

     This work has been supported in part by the grant-in-Aid for 
Scientific Research (05242102, 06233101, 4227) and COE research 
(07CE2002) of the Ministry of Education, Science, and Culture in Japan, 
and the Swiss Nationalfonds.

{\small
\begin{table}[h]
\caption{Nucleosynthesis products of SN II and Ia}
\label{tabmas}
\begin{tabular}{lcccccc}
\hline \hline
& \multicolumn{6}{c}{Synthesized mass ($M_\odot$)} \\
\cline{2-7} 
& \multicolumn{1}{c}{Type II} & \multicolumn{5}{c}{Type Ia} \\
Species & 10$\sim$50M$_\odot$ & W70 & W7 & WDD1 & WDD2 & WDD3  \\
\hline 
$^{12}$C  & 7.93E-02 & 5.08E-02 &4.83E-02 &1.80E-03 &1.15E-03 &5.82E-04 \\
$^{13}$C  & 3.80E-09 & 1.56E-09 &1.40E-06 &3.52E-08 &1.04E-08 &6.90E-09 \\
$^{14}$N  & 1.56E-03 & 3.31E-08 &1.16E-06 &3.08E-04 &2.44E-04 &7.42E-05 \\
$^{15}$N  & 1.66E-08 & 4.13E-07 &1.32E-09 &6.84E-07 &2.75E-07 &1.68E-07 \\
$^{16}$O  & 1.80     & 1.33E-01 &1.43E-01 &9.96E-02 &6.93E-02 &4.69E-02 \\
$^{17}$O  & 9.88E-08 & 3.33E-10 &3.54E-08 &3.99E-06 &4.50E-06 &1.72E-06 \\
$^{18}$O  & 4.61E-03 & 2.69E-10 &8.25E-10 &6.96E-07 &4.62E-07 &2.42E-07 \\
$^{19}$F  & 1.16E-09 & 1.37E-10 &5.67E-10 &1.68E-09 &8.90E-10 &5.48E-10 \\
$^{20}$Ne & 2.12E-01 & 2.29E-03 &2.02E-03 &1.45E-03 &9.13E-04 &6.77E-04 \\
$^{21}$Ne & 1.08E-03 & 2.81E-08 &8.46E-06 &4.09E-06 &1.47E-06 &2.30E-06 \\
$^{22}$Ne & 1.83E-02 & 2.15E-08 &2.49E-03 &1.34E-05 &1.96E-06 &1.39E-06 \\
$^{23}$Na & 6.51E-03 & 1.41E-05 &6.32E-05 &3.20E-05 &1.30E-05 &6.53E-06 \\
$^{24}$Mg & 8.83E-02 & 1.58E-02 &8.50E-03 &8.29E-03 &4.76E-03 &2.93E-03 \\
$^{25}$Mg & 1.44E-02 & 1.64E-07 &4.05E-05 &4.60E-05 &2.39E-05 &1.44E-05 \\
$^{26}$Mg & 2.01E-02 & 1.87E-07 &3.18E-05 &5.52E-05 &3.57E-05 &1.09E-05 \\
$^{27}$Al & 1.48E-02 & 1.13E-04 &9.86E-04 &4.65E-04 &2.74E-04 &1.73E-04 \\
$^{28}$Si & 1.05E-01 & 1.38E-01 &1.50E-01 &3.48E-01 &2.71E-01 &2.04E-01 \\
$^{29}$Si & 8.99E-03 & 6.03E-05 &8.61E-04 &6.05E-04 &3.87E-04 &2.49E-04 \\
$^{30}$Si & 8.05E-03 & 3.09E-05 &1.74E-03 &1.07E-03 &6.35E-04 &3.94E-04 \\
$^{31}$P  & 1.21E-03 & 8.51E-05 &4.18E-04 &2.67E-04 &1.80E-04 &1.23E-04 \\
$^{32}$S  & 3.84E-02 & 9.19E-02 &8.41E-02 &2.09E-01 &1.65E-01 &1.24E-01 \\
$^{33}$S  & 1.78E-04 & 5.83E-05 &4.50E-04 &3.48E-04 &2.49E-04 &1.74E-04 \\
\hline
\end{tabular}
\end{table}
\begin{table}[h]
\begin{tabular}{lcccccc}
\hline \hline
& \multicolumn{6}{c}{Synthesized mass ($M_\odot$)} \\
\cline{2-7} 
& \multicolumn{1}{c}{Type II} & \multicolumn{5}{c}{Type Ia} \\
Species & 10$\sim$50M$_\odot$ & W70 & W7 & WDD1 & WDD2 & WDD3  \\
\hline 
$^{34}$S  & 2.62E-03 & 2.84E-06 &1.90E-03 &3.42E-03 &2.50E-03 &1.75E-03 \\
$^{36}$S  & 1.78E-06 & 1.09E-11 &3.15E-07 &2.29E-07 &1.33E-07 &8.58E-08 \\
$^{35}$Cl & 1.01E-04 & 8.06E-06 &1.34E-04 &1.21E-04 &9.83E-05 &7.67E-05 \\
$^{37}$Cl & 1.88E-05 & 5.36E-06 &3.98E-05 &4.23E-05 &3.36E-05 &2.52E-05 \\
$^{36}$Ar & 6.62E-03 & 1.99E-02 &1.49E-02 &4.12E-02 &3.35E-02 &2.50E-02 \\
$^{38}$Ar & 1.37E-03 & 5.93E-07 &1.06E-03 &1.71E-03 &1.45E-03 &1.12E-03 \\
$^{40}$Ar & 2.27E-08 & 1.14E-12 &1.26E-08 &8.16E-09 &6.08E-09 &4.53E-09 \\
$^{39}$K  & 6.23E-05 & 1.82E-06 &8.52E-05 &1.05E-04 &9.00E-05 &7.00E-05 \\
$^{41}$K  & 5.07E-06 & 5.33E-07 &7.44E-06 &8.39E-06 &7.12E-06 &5.53E-06 \\
$^{40}$Ca & 5.77E-03 & 1.95E-02 &1.23E-02 &4.07E-02 &3.45E-02 &2.58E-02 \\
$^{42}$Ca & 4.23E-05 & 1.55E-08 &3.52E-05 &4.69E-05 &4.06E-05 &3.24E-05 \\
$^{43}$Ca & 1.08E-06 & 6.81E-08 &1.03E-07 &6.97E-08 &6.31E-08 &4.73E-08 \\
$^{44}$Ca & 5.53E-05 & 1.55E-05 &8.86E-06 &2.06E-05 &2.07E-05 &1.42E-05 \\
$^{46}$Ca & 1.43E-10 & 5.88E-11 &1.99E-09 &6.17E-08 &7.18E-08 &3.00E-08 \\
$^{48}$Ca & 5.33E-14 & 5.96E-12 &7.10E-12 &4.21E-06 &4.41E-06 &3.62E-06 \\
$^{45}$Sc & 2.29E-07 & 3.92E-08 &2.47E-07 &3.54E-07 &3.23E-07 &2.51E-07 \\
$^{46}$Ti & 7.48E-06 & 7.87E-07 &1.71E-05 &2.06E-05 &1.76E-05 &1.44E-05 \\
$^{47}$Ti & 2.11E-06 & 7.49E-07 &6.04E-07 &1.21E-06 &1.24E-06 &7.40E-07 \\
$^{48}$Ti & 1.16E-04 & 3.21E-04 &2.03E-04 &8.17E-04 &8.53E-04 &6.76E-04 \\
$^{49}$Ti & 5.98E-06 & 1.64E-06 &1.69E-05 &6.55E-05 &6.71E-05 &5.24E-05 \\
$^{50}$Ti & 3.81E-10 & 1.14E-05 &1.26E-05 &4.57E-04 &4.60E-04 &4.80E-04 \\
$^{50}$V  & 7.25E-10 & 5.66E-09 &8.28E-09 &5.49E-08 &5.69E-08 &4.52E-08 \\
$^{51}$V  & 1.00E-05 & 2.10E-05 &5.15E-05 &2.92E-04 &3.18E-04 &2.10E-04 \\
$^{50}$Cr & 4.64E-05 & 7.61E-05 &2.71E-04 &5.86E-04 &5.24E-04 &3.98E-04 \\
$^{52}$Cr & 1.15E-03 & 6.63E-03 &5.15E-03 &1.75E-02 &2.01E-02 &1.67E-02 \\
$^{53}$Cr & 1.19E-04 & 4.65E-04 &7.85E-04 &2.02E-03 &2.26E-03 &1.87E-03 \\
$^{54}$Cr & 2.33E-08 & 1.79E-04 &1.90E-04 &2.03E-03 &2.03E-03 &2.08E-03 \\
$^{55}$Mn & 3.86E-04 & 6.27E-03 &8.23E-03 &1.57E-02 &1.88E-02 &1.44E-02 \\
$^{54}$Fe & 3.62E-03 & 8.18E-02 &1.04E-01 &7.26E-02 &7.08E-02 &6.29E-02 \\
$^{56}$Fe & 8.44E-02 & 6.72E-01 &6.13E-01 &4.56E-01 &6.15E-01 &7.62E-01 \\
$^{57}$Fe & 2.72E-03 & 1.98E-02 &2.55E-02 &1.03E-02 &1.39E-02 &1.94E-02 \\
$^{58}$Fe & 7.22E-09 & 9.34E-04 &9.63E-04 &4.09E-03 &4.06E-03 &4.10E-03 \\
$^{59}$Co & 7.27E-05 & 6.28E-08 &1.02E-03 &6.83E-04 &8.60E-04 &1.12E-03 \\
$^{58}$Ni & 3.63E-03 & 9.67E-02 &1.28E-01 &3.01E-02 &3.34E-02 &4.62E-02 \\
$^{60}$Ni & 1.75E-03 & 1.43E-02 &1.05E-02 &4.03E-03 &4.15E-03 &4.78E-03 \\
$^{61}$Ni & 8.35E-05 & 2.30E-04 &2.51E-04 &7.93E-05 &9.23E-05 &1.17E-04 \\
$^{62}$Ni & 5.09E-04 & 1.37E-03 &2.66E-03 &1.26E-03 &1.36E-03 &1.88E-03 \\
$^{64}$Ni & 3.20E-14 & 1.22E-06 &1.31E-06 &3.45E-04 &3.35E-04 &3.61E-04 \\
$^{63}$Cu & 8.37E-07 & 2.86E-06 &1.79E-06 &3.91E-05 &4.25E-05 &3.49E-05 \\
$^{65}$Cu & 4.07E-07 & 9.60E-07 &6.83E-07 &1.48E-05 &1.59E-05 &1.00E-05 \\
$^{64}$Zn & 1.03E-05 & 1.05E-04 &1.22E-05 &3.54E-07 &6.97E-07 &2.56E-06 \\
$^{66}$Zn & 8.61E-06 & 8.65E-06 &2.12E-05 &3.15E-05 &3.18E-05 &3.74E-05 \\
$^{67}$Zn & 1.52E-08 & 3.96E-09 &1.34E-08 &3.55E-06 &3.90E-06 &1.98E-06 \\
$^{68}$Zn & 3.92E-09 & 3.29E-09 &1.02E-08 &4.38E-07 &4.94E-07 &2.81E-07 \\
\hline
\end{tabular}
\end{table}
}

\end{document}